\documentclass[twocolumn]{IEEEtran}
\usepackage{amsmath,amsfonts,amssymb,bm,mathrsfs}
\usepackage{color}
\usepackage{comment}
\usepackage{cite}
\usepackage{graphicx}
\graphicspath{{./Figs/}}
\newcommand{\req}[1]{(\ref{#1})}
\ifx\undefined\unit\def\unit#1{\ensuremath{\mathrm{#1}}}\fi

\newcommand{\sx}{\mathsf{x}}
\newcommand{\bx}{\bm{\mathsf{x}}}
\newcommand{\sy}{\mathsf{y}}
\newcommand{\by}{\bm{\mathsf{y}}}

\def\pageheads{\small\sffamily E. Rubiola \&al.\hfill The $\Omega$ Counter, 
a Frequency Counter Based on the Linear Regression
\hfill\today\qquad}
\author{E. Rubiola, M. Lenczner, P.-Y. Bourgeois, and F. Vernotte\thanks{%
Michel Lenczner, Pierre-Yves Bourgeois and Enrico Rubiola are with the CNRS
FEMTO-ST Institute, Department of Time and Frequency, UBFC, UFC, UTBM,
ENSMM, 26, chemin de l'\'Epitaphe, Besancon, France. E-mail \{michel.lenczner%
$|$pyb2$|$rubiola\}@femto-st.fr. Enrico's home page http://rubiola.org.}%
\thanks{%
F. Vernotte is with UTINAM, Observatory THETA of Franche-Comt\'{e},
University of Franche-Comt\'{e}/UBFC/CNRS, 41 bis avenue de l'observatoire - B.P.
1615, 25010 Besan\c{c}on Cedex, France (Email francois.vernotte@obs-besancon.fr).}}
\title{The $\Omega$ counter, a frequency counter based on the Linear Regression}

\begin{document}
\maketitle

\begin{abstract}
This article introduces the \boldmath$\Omega$ counter, a frequency counter --- or a frequency-to-digital converter, in a different jargon --- based on the Linear Regression (LR) algorithm on time stamps.

We discuss the noise of the electronics. We derive the statistical properties of the $\Omega$ counter on rigorous mathematical basis, including the weighted measure and the frequency response.  We describe an implementation based on a SoC, under test in our laboratory,  and we compare the $\Omega$ counter to the traditional $\Pi$ and $\Lambda$ counters.

The LR exhibits optimum rejection of white phase noise, superior to that of the $\Pi$ and $\Lambda$ counters.  White noise is the major practical problem of wideband digital electronics, both in the instrument internal circuits and in the fast processes which we may want to measure.

The $\Omega$ counter finds a natural application in the measurement of the Parabolic Variance, described in the companion article arXiv:1506.00687 [physics.data-an].
\end{abstract}

\section{Introduction and State of the Art}
The frequency counter in an instrument that measure the frequency of the input signal versus a reference oscillator.  Since frequency and time intervals are the most precisely measured physical quantity, and nowadays even fairly sophisticated counters fit in a small area of a chip, converting a physical quantity to a frequency is a preferred approach to the design of electronic instruments. The consequence is that the counter is now such a versatile and ubiquitous instrument that it \emph{creates applications} rather than being developed \emph{for applications}, just like the computer.

The term `counter' comes from the early instruments, where the Dekatron \cite{Dekatron}, a dedicated cold-cathode vacuum tube, was used to \emph{count} the pulses of the input signal at in a reference time, say 0.1 s or 1 s.
While the manufacturers of electrical instruments stick on the word `counter,' the new terms `Time-to-Digital Converter' (TDC) and `Frequency-to-Digital Converter' are generally preferred in digital electronics (see for example \cite{Henzler-2010:Time-to-digital-converters,yoder:quantizers}.

The direct frequency counter has long time ago been replaced by the classical reciprocal frequency counter, which measures the average period on a suitable interval by counting the pulses of the reference clock.  The obvious advantage is that the `$1/n$ counts' quantization uncertainty is limited by the clock frequency, instead of the arbitrary input frequency.  Of course, the clock is set by design at the maximum frequency allowed for the technology.

Higher resolution is obtained by measuring fractions of the clock period with a suitable interpolator.  Simple and precise interpolators work only at fixed frequency.  The most widely used techniques are described underneath.  Surprisingly, all them are rather old and feature picosecond range resolution.  The progress concerns the sampling rate, from kS/s or less in the early time to a few MS/s available now.  See \cite{Kalisz-2004-Metrologia--Counters} for a review, and \cite{Henzler-2010:Time-to-digital-converters} for integrated electronics techniques.
\begin{itemize}
\item The \emph{Nutt Interpolator} \cite{Nutt-1968-RSI--Counters,Nutt-1976-Patent--Interpolator} makes use of the linear charge and discharge of a capacitor.
\item The \emph{Frequency Vernier} is the electronic version of the `Vernier caliper' commonly used in the machine shop.  A synchronized oscillator close to the clock frequency plays the role of the Vernier scale \cite{Cottini-Gatti-1956-NC--Vernier,Cottini-Gatti-Giannelli-1956-NC--Vernier,Lefevre-1957:Vernier-chronotron,Kindlmann-Sunderland-RSI-1966--Chronotron}
\item The \emph{Thermometer Code Interpolator} uses a pipeline of small delay units and D-type flip-flops or latches.  To our knowledge, it first appeared in the HP\,5371A time interval analyzer implemented with discrete delay lines and sparse logic\cite{Chu-1989-HPJ--HP5371}.
\end{itemize}
Table~\ref{tab:tdcs} provide some examples of commercial products.

\begin{table}
\caption{Some commercial TDCs and Time Interval Analyzers.}
\label{tab:tdcs}
\sffamily\centering
\def\vrr{\vrule width 0.pt height 2.5ex depth 1.0ex}
\begin{tabular}{lclc}\hline
Device / Brand & Resolution & Type & Reference\vrr\\\hline
Acam TDC-GPX    & 10 ps & Chip & \cite{Acam}\vrr\\\hline
Brilliant                  & $<1$ ps & PCI/PXI Card & \cite{brilliant}\vrr\\\hline
Guidetech              & 1 ps & PCI/PXI Card & \cite{guidetech}\vrr\\\hline
K\&K                      & 50 ps & Special purpose & \cite{Kramer2001,K-K-Messtechnik}\vrr\\\hline
Keysight 53230A   & 20 ps & Lab instrument \ & \cite{Keysight}\vrr\\\hline
Maxim MAX35101 & 20 ps & Chip & \cite{Maxim}\vrr\\\hline
SPAD Lab TDC      & 22 ps & Chip & \cite{Spad}\vrr\\\hline
Texas THS788       & 8 ps   & Chip & \cite{texas}\vrr\\\hline
\end{tabular}
\end{table}

In the classical reciprocal counter, the input signal is sampled only at the beginning and at the end of the measurement time $\tau$.  There results a uniformly weighted averaging of frequency.  In the presence of time jitter of variance $\sigma^2_\sx$ (classical variance), statistically independent a start and stop, the  fluctuation of the measured fractional frequency has variance $\sigma^2_\sy=2\sigma^2_\sx/\tau^2$.
Increased resolution can be achieved with fast sampling and statistics.  A sampling interval $\tau_0=\tau/m$, $m\gg1$ enables averaging on $m$ highly overlapped and statistically independent measures.  In this way, one can expect a variance $\sigma^2_\sy\propto\sigma^2_\sx/m\tau^2$.  There results a triangle weighted averaging of frequency.  These two methods are referred to as $\Pi$ and $\Lambda$ counters (or estimators) because of the graphical analogy of the Greek letter to the weight function \cite{rubiola2005,dawkins2007}
The $\Pi$ and $\Lambda$ counters are related to the Allan variance AVAR \cite{allan1966,Barnes-et-al-1971-TIM--Frequency-stability} and to the Modified Allan variance MVAR \cite{Snyder-1980-AO--Mod-Allan,Allan-Barnes-1981-FCS--Mod-Allan,Lesage-Ayi-1984-IM--Mod-Allan}.  Frequency counters specialized for MVAR are available as a niche product, chiefly for research labs (for example \cite{Kramer2001}).

Having access to fast time stamping and to sufficient computing power on FPGA and SoC at low cost and acceptable complexity, we tackle the problem of the \emph{best} estimator to be implemented in a frequency counter.

The Linear Regression (LR) turns out to be the right answer because it provides the lowest-energy (or lowest-power) fit of a data set, which is the optimum approximation for white noise.
The LR can be interpreted as a weight function applied to the measured frequency fluctuations. The shape of such weight function is parabolic (Fig.~\ref{fig:weight}). We call the corresponding instrument `$\Omega$ counter,' for the graphical analogy of the parabola with the Greek letter, and in the continuity of the $\Pi$ and $\Lambda$ counters \cite{rubiola2005, dawkins2007}. The $\Omega$ estimator is similar to the $\Lambda$ estimator, but exhibits higher rejection of the instrument noise, chiefly of white phase noise. This is important in the measurement of fast phenomena, where the noise bandwidth is necessarily high, so the white phase noise is integrated over the bandwidth.

The idea of the LR for the estimation of frequency is not new \cite{Chu-1989-HPJ--HP5371,Johansson2005fcs-counters}.  However, these articles lack rigorous statistical analysis.  Another modern use of the LR has been proposed independently by Benkler et al.\ \cite{Benkler2015arxiv} at the IFCS, where we gave our first presentation on the $\Omega$ counter and on our standpoint about the Parabolic Variance PVAR\@.

\section{Instrument Architecture and Noise}
\begin{figure}
\centering\includegraphics[width=84mm]{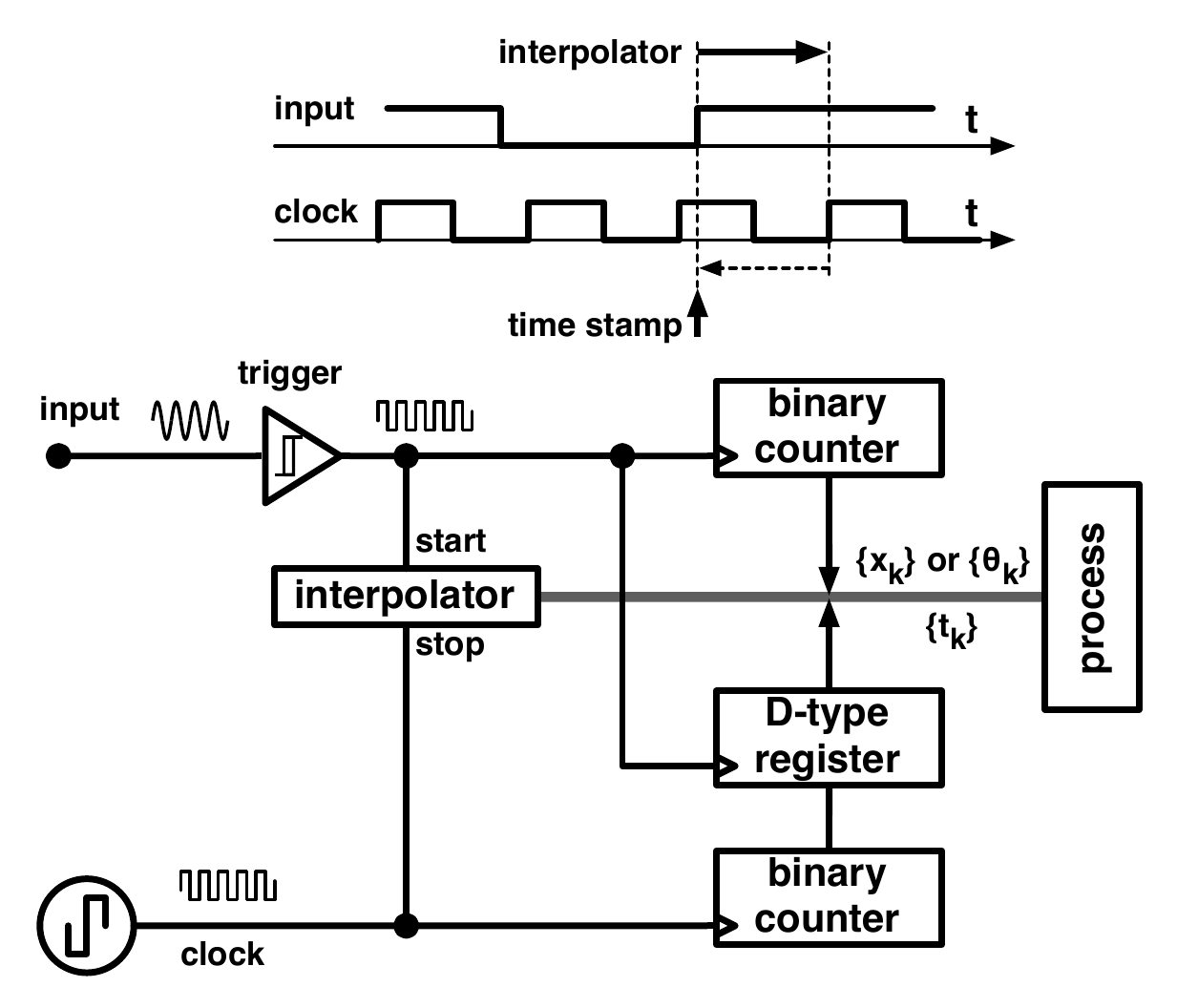}\par
\caption{Time-stamp counter architecture.}
\label{fig:architecture}
\end{figure}
The use of time stamps for sophisticated statistics is inspired to the `Picket Fence' method introduced by C. Greenhall for the JPL time scale \cite{Greenhall-1989-UFFC--Picket-fence,Greenhall-1997-IM--Picket-fence}.
Figure~\ref{fig:architecture} shows a rather general block diagram.
Following the signal paths, we expect that all the noise originating inside the instrument contains only white and flicker PM noise.  The reason is that the instrument internal delay cannot diverge in the long run.  Notice that the divergence of the flicker is only an academic issue \cite{Vernotte-2015-Metrologia}.  Integrating over 20--40 decades of frequency, the rms delay exceeds by a mere 15--20 dB the flicker coefficient.  Conversely the reference oscillator, which in a strict sense are not a part of the instrument, and the signal under test can include white FM and slower phenomena.   The analysis of practical cases, underneath, shows that the instrument internal noise is chiefly white PM\@.

\subsection{Internal clock distribution}
The reference clock signal is distributed to the critical parts of the counters by appropriate circuits.  As an example, we evaluate the the jitter of the Cyclone III FPGA by Altera \cite{Altera}.
A reason is that we have studied thoroughly the noise of this device \cite{Calosso-2014-IFCS--Digital}.  Another reason is that it is representative of the class of mid sized FPGAs, and similar devices from other brands, chiefly Xilinx\cite{Xilinx}, would give similar results.

White phase noise is of the aliased $\varphi$-type, described by
\begin{align*}
S_\varphi(f)=\frac{k^2}{\nu_0}
\end{align*}
where $\nu_0$ is the clock frequency, and $k\approx$ 630 $\mu$rad is an experimental parameter of the component.
Since the PM noise is sampled at the threshold crossings, i.e., at $2\nu_0$, the  bandwidth is equal to $\nu_0$.
Converting $S_\varphi(f)$ into $S_\sx(f)=\frac{1}{4\pi^2\nu_0^2}S_\varphi(f)$ and integrating on $\nu_0$, we get
\begin{align*}
\left<\sx^2\right>=\frac{k^2}{4\pi^2\nu_0^2}
\end{align*}
thus $\sx_\text{rms}=2.8$ ps at 100 MHz clock frequency.

Flicker phase noise is of the $\sx$-type.  We measured a value of 22 \unit{fs/\sqrt{Hz}} at 1 Hz, and referred to 1 Hz bandwidth.  Integrating over 12-15 decades, we find 115--130 fs rms, which is low as compared to white noise.

\subsection{Input trigger}
In most practical cases the noise is low enough to avoid multiple bounces at the threshold crossings, as described in \cite{Rubiola-1992-FCS-Trigger}
In this condition, the rms time jitter is
\begin{align*}
\sx_\text{rms}=\frac{V_\text{rms}}{\text{SR}}
\end{align*}
where $V_\text{rms}$ is the rms fluctuation of the threshold, and SR is the slew rate (slope, not accounting for noise).

The rms voltage results from white and flicker noise.  The former is described as $V_w=e_n\sqrt{B}$, where $e_n$ is the white noise in 1 Hz bandwidth, and $B$ the noise bandwidth.  The latter results from $V_f=\alpha e_n\sqrt{\ln(B/A)}$, where $\alpha e_n$ is the flicker noise at 1 Hz and referred to 1 Hz bandwidth, and $A$ is the low cutoff frequency.

The golden rules for precision high-speed design suggest
\begin{itemize}
\item White noise, $e_n=10$ \unit{nV/\sqrt{Hz}}, including the input protection circuits (without, it would be of 1--2 \unit{nV/\sqrt{Hz}}).
\item Flicker noise, $\alpha=10\ldots31.6$ (20--30 dB) at most, as a conservative estimate
\item the noise bandwidth is related to the maximum input (toggling) frequency $\nu_\text{max}$ by $\nu_\text{max}=0.3\,B$.
\end{itemize}

Let us consider a realistic example where $\nu_\text{max}=1.2$ GHz, $e_n=10$ \unit{nV/\sqrt{Hz}}, and $\alpha=31.6$ (30 dB).  Accordingly, the input bandwidth is $B=4$ GHz.  In turn the integrated white noise is $V_w=632$ $\mu$V rms.  By contrast, flicker is $V_f=1.66$ $\mu$V rms
integrated over 12 decades (4 mHz to 4 GHz), and $V_f=1.86$ $\mu$V rms integrated over 15 decades (4 $\mu$Hz to 4 GHz).  White noise is clearly the dominant effect.

\subsection{Clock interpolator}
Commercially available counters exhibit single-shot fluctuation of 1--50 ps [Tab.~\ref{tab:tdcs} and References herein].  Since the interpolator is reset to its initial state after each use, it is sound to assume that the noise realizations are statistically independent, which is white noise.  At a closer sight some memory between operations is possible, which shows up as some flicker noise.  However, the same issues about bandwidth already discussed in this Section apply, and we expect that flicker is a minor noise effect as compared to white noise.

\subsection{Motivation for the Linear Regression}
The LR finds a straightforward application to the estimation of the frequency $\nu$ of a periodic phenomenon $\sin[\bm\phi(t)]$ from its phase $\bm\phi(t)$ using the trite relation $\nu=\frac{1}{2\pi}\,\frac{d\bm\phi}{dt}$.
It is well known that the LR provides the best estimate $\hat{\nu}$ in the presence of white noise.  The term `best' means that the LR minimizes the squared residuals.  This property matches the need of rejecting the white phase noise, which is the the main concern in counters, as we will see underneath.

\section{The Linear Regression}
\noindent
A real sinusoidal signal affected by noise can be written as
\begin{align*}
v(t)=V_0\sin[\bm\phi(t)]
\end{align*}
where $V_0$ is the amplitude, $\nu_0$ is the nominal frequency, and $\bm\phi(t)$ is a phase that carries both the `ideal time' and randomness.  The randomness in $\bm\phi(t)$ can be interpreted as either a phase fluctuation or a frequency fluctuation
\begin{align*}
\bm\phi(t)=\begin{cases}
2\pi\nu_0t+\varphi(t) & \text{PM noise representation}\\
2\pi\nu_0t+\int(\Delta\nu)(t)\:dt & \text{FM noise representation}
\end{cases}
\end{align*}
Of course $\bm\phi(t)$ and $\varphi(t)$ are allowed to exceed $\pm\pi$.

\begin{figure}
\centering\includegraphics[width=80mm]{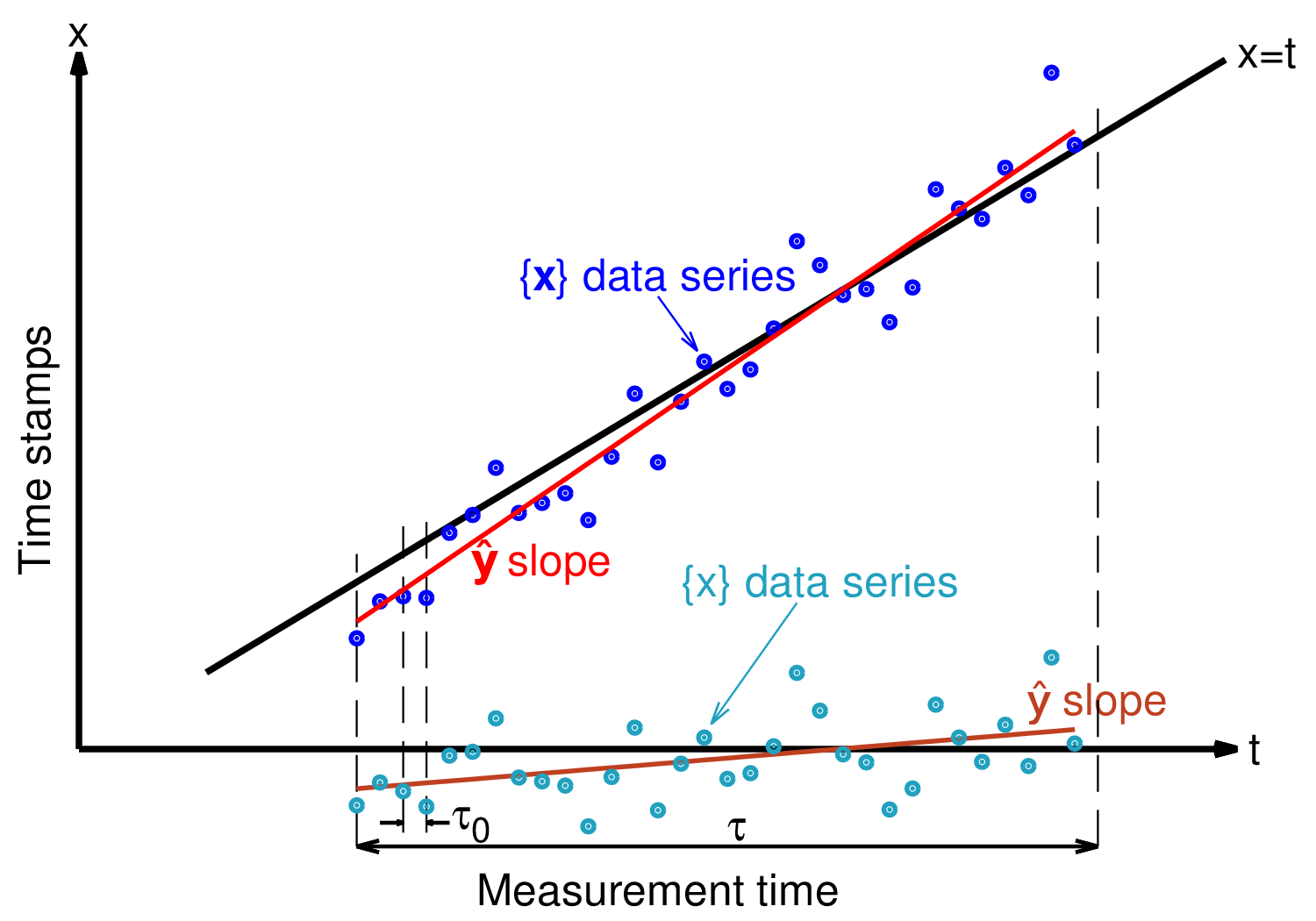}\par
\caption{Principle of the LR counter, and definition of often used variables.}
\label{fig:notation}
\end{figure}

We prefer to derive the properties of the LR using the normalized quantities $\bx(t)$ and $\by(t)$, as in Fig.~\ref{fig:notation}
\begin{align}
\bx(t)&=t+\sx(t)&&\text{(phase time)}\label{eqn:x-def}\\
\by(t)&=1+\sy(t)&&\text{(fractional frequency)}\label{eqn:y-def}
\intertext{where}
\sx(t)&=\varphi(t)/2\pi\nu_0\\
\by(t)&=\dot{\bx}(t).
\end{align}
The quantity $\bx(t)$ is the \emph{time} carried by the real signal, which is equal to the ideal time $t$ plus the random fluctuation $\sx(t)$.  The quantities $\sx(t)$ and $\sy(t)$ match the phase time fluctuation $x(t)$ and the fractional frequency fluctuation $y(t)$ used in the general literature \cite{allan1966,Barnes-et-al-1971-TIM--Frequency-stability}, with the choice of the font as the one and only difference.

Most concepts are suitable to continuous and the discrete treatise with simplified notation.   For example, $\bx=t+\sx$ maps into $\bx_{k}=t_{k}+\sx_{k}$ for sampled data, and into $\bx(t)=t+\sx(t)$ in the continuous case. The
notation $\left\langle\,.\,\right\rangle$ and $(.,.)$ stands for the average and the scalar product, defined as $\left<x\right>=\frac{1}{n}\sum_{k}x_{k}$ and
$(x,y)=\sum_{k}x_{k}y_{k}$ for time series, where $n$ is the number of terms in the sum, or as $\left<x\right>=\frac{1}{T}\int x(t)\,dt$ and $(x,y)=\int x(t)y(t)\,dt$, where $T$ is the integration time in the continuous case.  The span of the sum and the integral will be made precise in each case of application. The norm is defined as $||x||=\smash{\sqrt{(x,x)}}$.  The mathematical expectation and the variance of random variables are denoted by $\mathbb{E}\{\,.\,\}$ and $\mathbb{V}\{\,.\,\}$.

The problem of the linear regression consists in identifying the optimum value $\hat{\by}$ of the slope $\eta$ (dummy variable used to avoid confusion with $\by$) that minimizes the norm of the error $\bx-\eta t$. The solution is the random variable
\begin{equation}
\hat{\by}=\frac{\left( \bx-\left\langle \bx%
\right\rangle ,t-\left\langle t\right\rangle \right) }{||t-\left\langle
t\right\rangle ||^{2}}.
\label{eqn:LR-general}
\end{equation}
The choice of a reference system where the time sequence is centered at
zero, i.e.,  $\left\langle t\right\rangle =0$, makes the treatise simpler, without loss of generality.  Accordingly, the estimator $\hat{\by}$ reads
\begin{equation}
\hat{\by}=\frac{\left( \bx,t\right) }{||t||^{2}}.
\label{eqn:LR-centered}
\end{equation}
These choice are equivalent because $\left\langle t\right\rangle =0$, so $\left( \bx%
-\left\langle \bx\right\rangle ,t\right) =\left( \bx,t\right)
-\left\langle \bx\right\rangle \left( 1,t\right) =\left( \bx%
,t\right) $.

\section{Basic Statistical Properties}

\subsection{Unbiased estimate}
The LR provides an unbiased estimate of the slope, i.e.,
\begin{align*}
\mathbb{E}\{\hat{\by}\}&=1&&\text{(unbiased estimate)},
\end{align*}
even without need that the noise samples (or values) are
statistically independent. This is seen by replacing the expression
of the phase
\begin{equation*}
\hat{\by}=\frac{(\bx,t)}{||t||^{2}}=\frac{(t+\mathsf{%
x},t)}{||t||^{2}}=1+\frac{(\sx,t)}{||t||^{2}}
\end{equation*}%
so
\begin{equation*}
\mathbb{E}\{\hat{\by}\}=1+\frac{(\mathbb{E}\{\sx\},t)}{%
||t||^{2}}=1.
\end{equation*}

\subsection{Estimator variance}

Adding the hypothesis that samples $\sx_{k}$ (or the values $\mathsf{x%
}(t)$) are independent, then the estimator variance is
\begin{align*}
\mathbb{V}\{\hat{\by}\}&=\frac{\sigma _{\sx}^{2}}{||t||^{2}%
}&&\text{(estimator variance).}
\end{align*}%
This is seen by expanding the variance%
\begin{eqnarray*}
\mathbb{V}\{\hat{\by}\} &=&\mathbb{E}\{(\hat{\by}-%
\mathbb{E}\{\hat{\by}\})^{2}\} \\
&=&\mathbb{E}\{\hat{\by}^{2}\}-\mathbb{E}\{\hat{\by}%
\}^{2}.
\end{eqnarray*}
But
\begin{align*}
\mathbb{E\{(}\hat{\by})^{2}\}& =\frac{1}{||t||^{4}}\mathbb{E}%
\{(t+\sx,t)^{2}\} \\
& =\frac{1}{||t||^{4}}\mathbb{E}\{(t,t)^{2}+2(t,t)(\sx,t)+(\sx%
,t)^{2}\} \\
& =1+\frac{1}{||t||^{4}}\mathbb{E}\{(\sx,t)^{2}\}.
\end{align*}%
Since the random variables $\sx_{k}$ are independent, then
\begin{align*}
\mathbb{E}\{(\sx,t)^{2}\}& =\sum_{k,\ell }t_{k}t_{\ell }\:\mathbb{E}\{%
\sx_{k}\sx_{\ell }\} \\
& =\sum_{k}t_{k}^{2}\:\mathbb{E}\{\sx_{k}^{2}\} \\
& =||t||^{2}\sigma _{\sx}^{2}.
\end{align*}%
We conclude
\begin{equation*}
\mathbb{V}\{\hat{\by}\}=1+\frac{||t||^{2}}{||t||^{4}}\sigma _{%
\sx}^{2}-1=\frac{\sigma _{\sx}^{2}}{||t||^{2}}.
\end{equation*}

\subsection{Uniformly spaced time series}

In the case of a uniformly spaced data series with a constant step $\tau _{0}
$ and with $\tau =m\tau _{0}$, then for large $m$ it holds that
\begin{gather}
\hat{\by}\approx 1+\frac{12(\sx,t)}{m\tau ^{2}} \nonumber\\[1ex]
\mathbb{V}\{\hat{\by}\}\approx \frac{12\sigma _{\sx}^{2}}{m\tau ^{2}}.
\label{eqn:Omega-variance}
\end{gather}
For even $m=2p$, the proof starts with $t_{k}=k\tau _{0}$ for $k\in \{-p,...,p\}$,%
\begin{eqnarray*}
||t||^{2} &=&\sum\limits_{k=-p}^{p}t_{k}^{2}=\tau
_{0}^{2}\sum\limits_{k=-p}^{p}k^{2}=2\tau _{0}^{2}\sum\limits_{k=1}^{p}k^{2}
\\
&=&\frac{\tau _{0}^{2}}{3}p\left( 2p+1\right) \left( p+1\right) =\frac{\tau
_{0}^{2}}{12}m\left( m+2\right) \left( m+1\right)  \\
&\approx &\frac{m\tau ^{2}}{12}\qquad \text{for large}~m.
\end{eqnarray*}%
A similar calculation can be carried out for odd $m=2p+1$ with $t_{k}=(k+%
\frac{1}{2})\tau _{0}$ for $k\in \{-p-1,...,p\}$.

\subsection{Continuous input signal}
In the case of input signal continuous over a symmetric time interval
$(-\frac{\tau }{2},\frac{\tau }{2})$, we get
\begin{equation*}
\hat{\by}=1+\frac{12(\sx,t)}{\tau ^{3}}\text{.}
\end{equation*}%
Evidently $\left\langle t\right\rangle =0$, so the first equality results
from $||t||^{2}=\tau ^{3}/12$. Moreover, for a white noise $\sx$
\begin{equation*}
\mathbb{V}\{\hat{\by}\}=\frac{12\sigma _{\sx%
}^{2}}{\tau ^{2}}.
\end{equation*}%
Indeed, replacing $\hat{\by}$ by its expression,
$$
\mathbb{V}\{\hat{\by}\}=\mathbb{V}\left\{1+\frac{(\sx,t)}{||t||^{2}}\right\}=%
\frac{1}{||t||^{4}}\mathbb{V}\{(\sx,t)\}.
$$
But $\mathbb{E}\{(\mathsf{x%
},t)\}=(\mathbb{E}\{\mathsf{x\}},t)=0$, so
\begin{align*}
\mathbb{V}\{\hat{\by}\}
&=\frac{1}{||t||^{4}}\mathbb{E}\{(\mathsf{x},t)^{2}\}\\
&=\frac{1}{||t||^{4}}\int_{-\tau /2}^{\tau /2}\int_{-\tau /2}^{\tau
/2}\mathbb{E}\{\sx(t)\mathsf{x(s)}\}\,ts\,dt\,ds.
\end{align*}%
Since $\mathbb{E}\{\sx(t)\mathsf{x(s)}\}=\sigma _{\sx%
}^{2}\delta (t-s)$,%
\begin{equation*}
\mathbb{V}\{\hat{\by}\}=\frac{\sigma _{\sx}^{2}}{||t||^{4}%
}\int_{-\tau /2}^{\tau /2}t^{2}\,dt=\frac{\sigma _{\sx}^{2}}{%
||t||^{2}}
\end{equation*}%
and the conclusion follows.

\section{Weighted Average and Filtering}
We show that the Omega counter can be described as a weighted average or as a filter.  The impulse response is derived from the weight function.

\subsection{Weighted measure}
Let us consider an instrument that measures $\by(t)$ averaged on time interval of duration $\tau$.  We express the estimate $\hat{\by}(t)$ available at the time $t$ as the weighted average
\begin{align}
\hat{\by}(t)&=\int_\mathbb{R}\by(s)\,w(s-t)\:ds,
\label{eqn:w-def}
\end{align}
where $w(t)$ is a suitable weight function which we will identify later in this Section, ruled by
\begin{align*}
&w(t)=0~~\text{for}~t\not\in\{-\tau,0\}    &&\text{(interval)}\\
&\int_\mathbb{R}w(t)\:dt=1        &&\text{(normalization)}.
\end{align*}
These rules expresses the fact that the integration time has duration $\tau$ ending at the time $t$, and that the normalization yields a valid average.

Eventually, \req{eqn:w-def} can be seen as a scalar product, or as an operator on $\by$, or as a weighted measure as regularly used in physics and engineering, or as a measure in the mathematical measure theory.

We first identify a function $\tilde{w}(t)$ which satisfies
\begin{align}
\hat{\by}(t)&=\int_\mathbb{R}\bx(s)\,\tilde{w}(s-t)\:ds.
\label{eqn:wtilde-def}
\end{align}
Then, remembering that $\by(t)=\dot{\bx}(t)$ and using the integration by part formula $\int f(t)g'(t)\,dt=-\int f'(t)g(t)\,dt$, we find
\begin{align}
w(t)&=-\int_{-\infty}^{t}\tilde{w}(s)\,ds.
\label{eqn:wtilde-integral}
\end{align}

Since we prefer the simplified expression \req{eqn:LR-centered} for the LR, we work in the $(-\frac{\tau }{2},\frac{\tau }{2})$ interval using the centered functions
\begin{align*}
w_c(t)&=w(t+\tau/2)&
\tilde{w}_c(t)&=\tilde{w}(t+\tau/2).
\end{align*}
Replacing $\tilde{w}(t)$ with $\tilde{w}_c(t)$, Eq.~\req{eqn:wtilde-def} gives the estimate at the time $t+\tau/2$
\begin{align}
\hat{\by}(\tau/2)&=\int_{-\tau/2}^{\tau/2}\bx(t)\,\tilde{w}_c(t)\:dt.
\label{eqn:y-wtilde-c}
\end{align}

The LR (Eq.~\req{eqn:LR-centered}) applied over $(-\frac{\tau }{2},\frac{\tau }{2})$ gives
\begin{align}
\hat{\by}(\tau/2)&=\frac{(\bx,t)}{||t||^2}=
\frac{12}{\tau^3}\int_{-\tau/2}^{\tau/2} \bx(t)\,t\, dt.
\label{eqn:y-LR}
\end{align}

\begin{figure}
\centering\includegraphics[width=72mm]{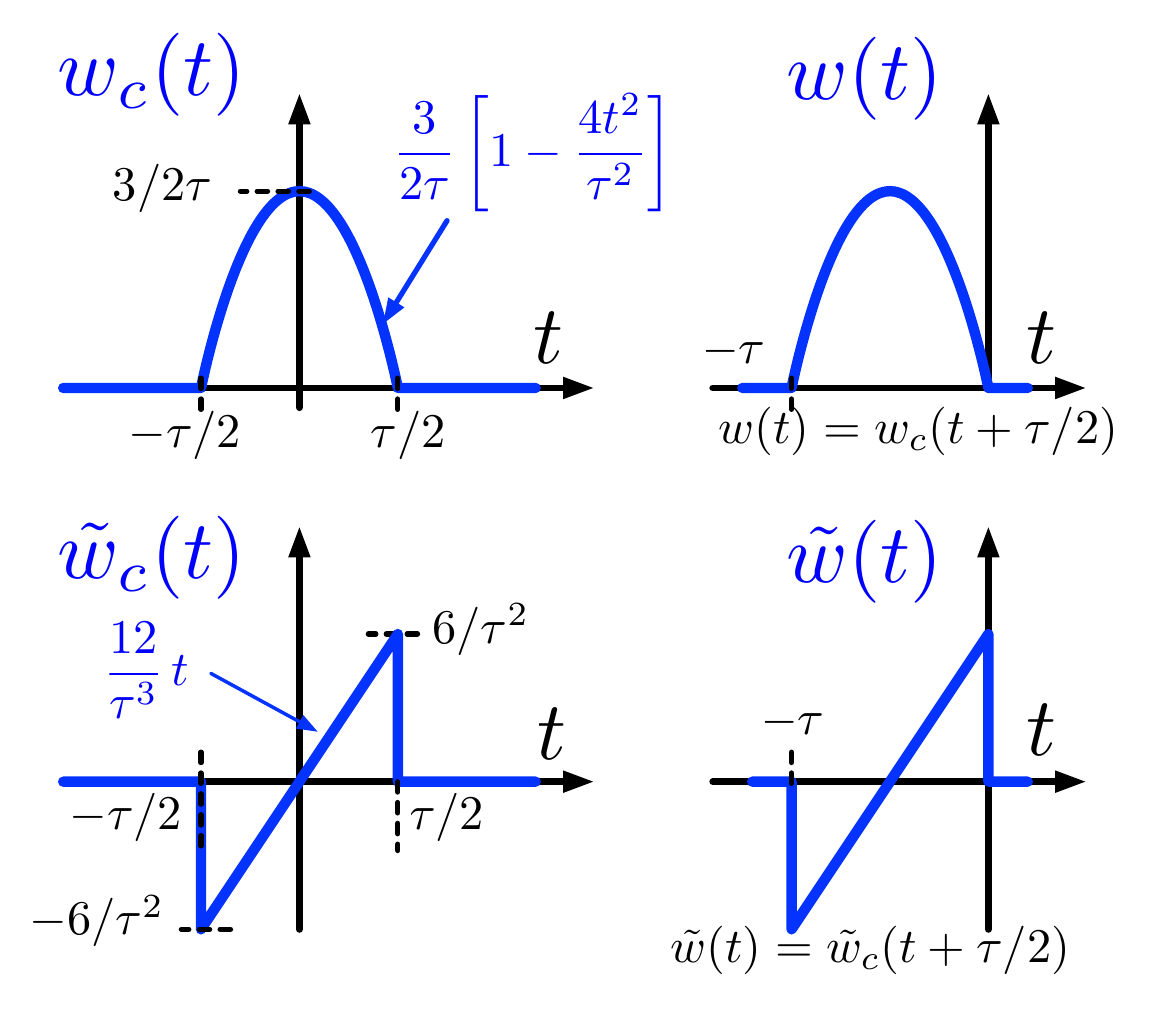}\par
\caption{Weight functions of the $\Omega$ counter.}
\label{fig:weight}
\end{figure}

Comparing \req{eqn:y-wtilde-c} and \req{eqn:y-LR}, we find
\begin{align}
\tilde{w}_c(t)=\frac{12}{\tau^3}\,t
\quad\text{for}~t\in\textstyle(-\frac{\tau }{2},\frac{\tau }{2}),
\quad0~\text{elsewhere},
\end{align}
and, using Eq.~\req{eqn:wtilde-integral},
\begin{align*}
w_c(t)=\frac{3}{2\tau}\left[1-\frac{4t^2}{\tau^2}\right]
\quad\text{for}~t\in\textstyle(-\frac{\tau }{2},\frac{\tau }{2}),
\quad0~\text{elsewhere}.
\end{align*}
These functions are plotted in Fig.~\ref{fig:weight}.

\subsection{Frequency response}
For the general experimentalist, the noise rejection properties of the counter are best seen in the frequency domain.
The fluctuations $\sy(t)$ and $\sx(t)$ are described in terms of their single-sided power spectral densities (PSD) $S_\sy(f)$ and $S_\sx(f)$, and the counter as a LTI (linear time-invariant) system which responds with its output fluctuation $\hat{\sy}(t)$.  The counter output PSD is
\begin{align*}
S_{\hat{\sy}}(f)&=|H(f)|^2S_{\sy}(f)\\
S_{\hat{\sy}}(f)&=|\tilde{H}(f)|^2S_{\sx}(f),
\end{align*}
where $|H(f)|^2$ and $|\tilde{H}(f)|^2$ are the frequency response for frequency noise and phase noise, respectively.

The LTI system theory teaches us that $H(f)$ is the Fourier transform of the impulse response $h(t)$
\begin{align*}
\tilde{H}(f) &=\int_{\mathbb{R}}\tilde{h}(t)e^{-2i\pi ft}\,dt,
\end{align*}
and so $\tilde{h}(t)\leftrightarrow\tilde{H}(f)$.

Notice that a time shift $t\rightarrow t\pm\frac{\tau}{2}$ introduces a term $e^{\pm i\pi\tau f}$ in the Fourier transform, which vanishes in $|H(f)|^2$ and $|\tilde{H}(f)|^2$.  So, we can use the centered version, $h_c(t)\leftrightarrow H_c(f)$ and $\tilde{h}_c(t)\leftrightarrow\tilde{H}_c(f)$, of the above functions.

\begin{figure}[t]
\centering\includegraphics[width=72mm]{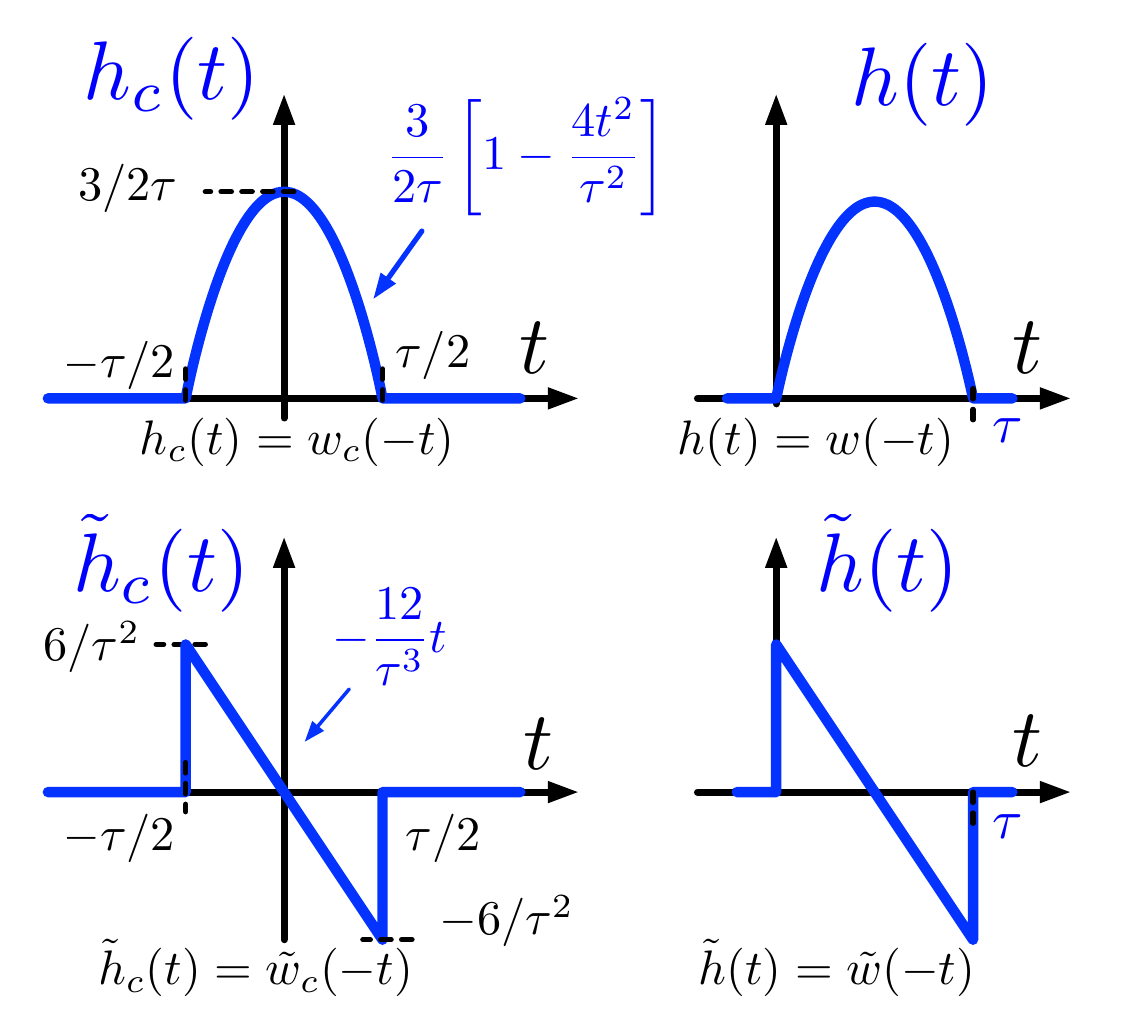}\par
\caption{Impulse response functions of the $\Omega$ counter.}
\label{fig:kernel}
\end{figure}

We start from the time-domain response of the estimator, which results from the convolution integral
\begin{align*}
\hat{\sy}(t)
&=\sy(t)\ast h(t)\\
&=\int_\mathbb{R}\sy(s)\,h(t-s)\,ds.
\end{align*}
A direct comparison of the above to \req{eqn:w-def} gives $h(t)=w(-t)$, hence (Fig.~\ref{fig:kernel})
\begin{align*}
h_c(t)=\frac{3}{2\tau}\left[1-\frac{4t^2}{\tau^2}\right]
\quad\text{for}~t\in\textstyle(-\frac{\tau }{2},\frac{\tau }{2}),
\quad0~\text{elsewhere},
\end{align*}
and
\begin{gather*}
\tilde{H}_c(f)=-\frac{6i}{\pi ^{2}f^{2}\tau ^{3}}\left( \pi f\tau \cos
(\pi f\tau )-\sin (\pi f\tau )\right)\\
|\widetilde{H}_c(f)|^{2}=\frac{36}{\pi ^{4}f^{4}\tau ^{6}}\left( \pi f\tau
\cos (\pi f\tau )-\sin (\pi f\tau )\right) ^{2}.
\end{gather*}

In the same way, comparing the convolution integral
\begin{align*}
\hat{\sy}(t)
&=\sx(t)\ast \tilde{h}(t)\\
&=\int_\mathbb{R}\sx(s)\,\tilde{h}(t-s)\,ds
\end{align*}
to \req{eqn:wtilde-def} yields $\tilde{h}(t)=\tilde{w}(-t)$.  There follows (Fig.~\ref{fig:kernel})
\begin{align}
\tilde{h}_c(t)=
-\frac{12}{\tau^3}\,t\quad\text{for}~t\in(\textstyle -\frac{\tau }{2},\frac{\tau }{2}),
\quad 0~\text{elsewhere}
\end{align}
and
\begin{gather*}
H_c(f)=-\frac{3}{\pi ^{3}f^{3}\tau ^{3}}\left( \pi f\tau \cos (\pi f\tau
)-\sin (\pi f\tau )\right)\\
|H_c(f)|^{2}=\frac{9}{\pi ^{6}f^{6}\tau ^{6}}\left( \pi f\tau \cos (\pi f\tau
)-\sin (\pi f\tau )\right) ^{2}.
\end{gather*}

\begin{figure}[t]
\includegraphics[width=80mm]{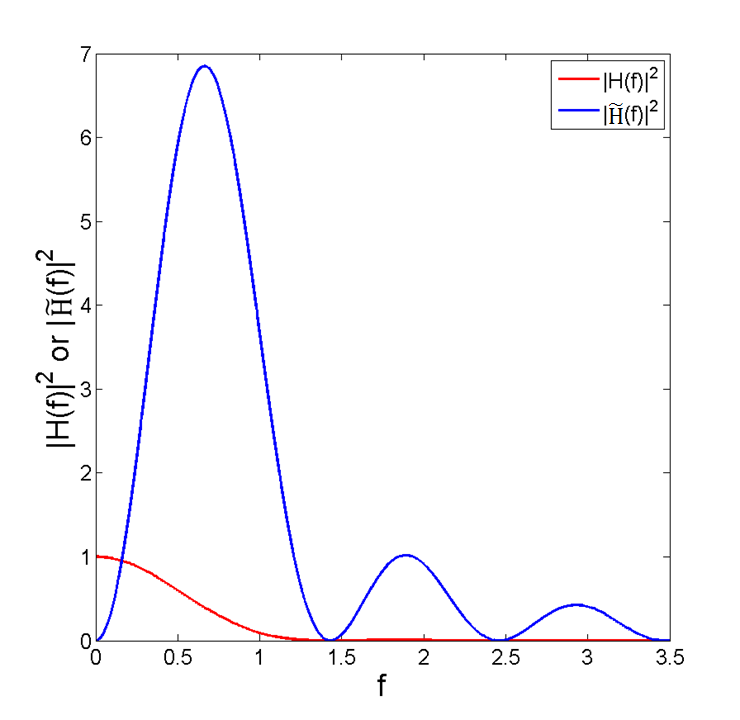}\par
\caption{Frequency response of the $\Omega$ counter, plotted for $\tau=1$ s..  $|H_c(f)|^{2}$ is the response to fractional-frequency noise $S_\sy(f)$, while $|\tilde{H}_c(f)|^{2}$ is the response to phase-time noise $S_\sx(f)$.}
\label{fig:frequency-response}
\end{figure}
The frequency transfer functions $|H(f)|^2$ and $|\tilde{H}(f)|^2$ are plotted in Fig.~\ref{fig:frequency-response}.

\section{Hardware Techniques}
We describe a hardware implementation.
Dropping the normalization, the LR estimates the frequency
\begin{equation}
\hat{\nu} =
\dfrac{\sum_{k=0}^{m-1}\left(\bm\theta_k-\left<\bm\theta_k\right>\right)\left(t_k-\left<t_k\right>\right)}{\sum_{k=0}^{m-1}\left(t_k-\left< t_k\right>\right)^2}
\end{equation}
by fitting the phase data $\bm\theta_k=\bm\phi_k/2\pi$ expressed as the fraction of period.

It is worth mentioning that timing indexes form a series of known natural terms such that the pre evaluation of $\{t_k\}$ and its average $\left<t_k\right>$ should be considered.  The result of the average depends on $m$, the number of iterations.  With $m$, some pre-calculated registers can be transferred into a RAM.

\begin{figure}
\centering\includegraphics[width=84mm]{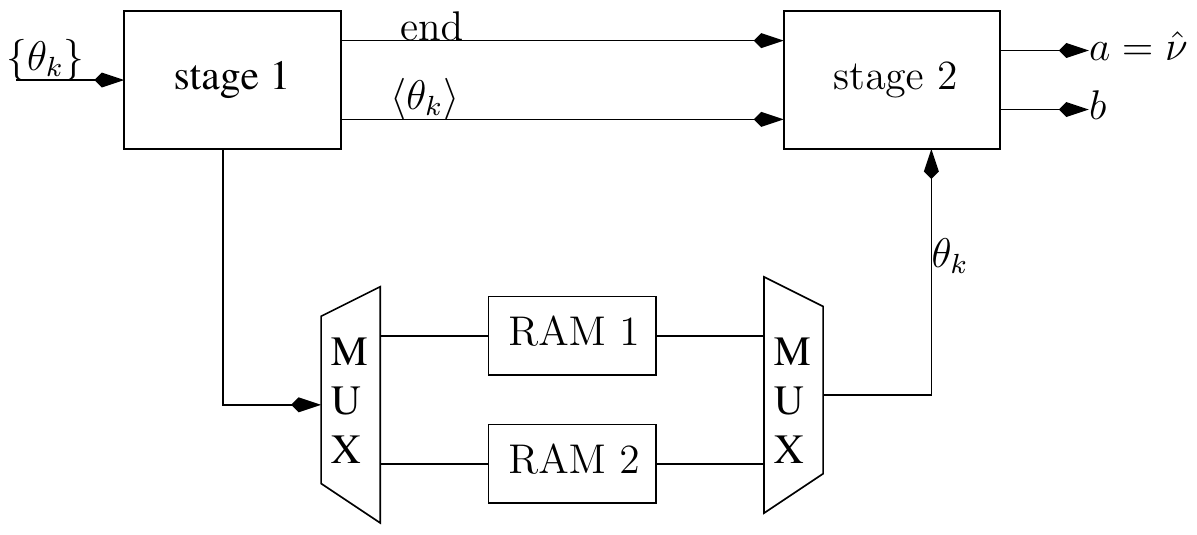}
\caption{Hardware implementation.}
\label{fig:Omega-block-diagram}
\end{figure}

Fig. \ref{fig:Omega-block-diagram} shows the block diagram.
The algorithm takes two steps, represented as `stages.'  The reason is that both average and the samples over $m$ are needed at the same time.
As a consequence, the first stage is mainly devoted to the evaluation of the average and the transfer of the data stream into an external dual-port RAM, the second stage to the calculation of the LR coefficients.

\subsection*{First stage}
\subsubsection{Data transfer}
The incoming data flow is transferred into an external RAM at the interpolator sampling rate $1/\tau_0$.
In the presented design, $1/\tau_0$ is the effective sample rate of the data stream.

\subsubsection{Accumulation}
Assuming $\{\theta_k\}$ where each $\theta_k$ is coded in $M$ bits, the data stream is firstly
extended to $2M$ bits and shifted by a selected amount of gain $g=\log_2(m)$.
This way, when the accumulation process is done, there is no need to divide the result to obtain the
average, by getting rid of the $g$ lower bits. This process preserves the data integrity as the gain
satisfies $g\leqslant M$.

When the average is calculated and the first RAM full, an `end' of process signal and the result are propagated into the second stage.
A new process cycle starts with the next data stream that will be stored in the second RAM. Such a cycle lasts obviously $\tau=m\tau_0$.
The second stage is running at the same time, with the same duration $\tau$.
Apart from an initial time lag $\tau$,
this mechanism enable to calculate the LR coefficients at the rate of $1/\tau_0$ (so with no data loss): stage 2 is calculating on the old $m$ data while the first calculates on the newer data.

\subsection*{Second stage}
Here the averages are known and the samples collected, the calculation of the LR parameters are performed within this block.

\subsubsection{Process 1}Differences calculations.
The samples are collected from the RAM and $\bm\theta_i-\left<\bm\theta_k\right>$ is calculated.
In parallel, $t_i-\left<t_k\right>$ are also computed, with $\left<t_k\right>$ provided by the CPU.
The calculation of $t_i-\left<t_k\right>$ is cost effective compared to the RAM resource usage.
All the data are latched.

\subsubsection{Process 2} frequency estimation.
This process is an asynchronous accumulation $\sum_n(\bm\theta_k-\left<\bm\theta_k\right>)(t_k-\left<t_k\right>)$.
The slope $a$ is obtained by dividing the preceeding result by the constant denominator term provided by the CPU and stored into a register, and $\hat{\nu}$ is latched.  Then, the estimated parameters are resized and propagated.

\subsection*{Oracle}
A testbench of the design requires different data sets to correctly predict its behavior.
For we have implemented a C-based version code of the design in fixed point arithmetics, and equivalent to the GNU-Octave {\it{LinearRegression}} function or the Levenberg-Marquardt algorithm (as embedded into Gnuplot for example).

A hardware implementation works at a rate of $1/\tau_0=250$ MHz.

The chip area occupation is only limited by the number of iterations $m$ which
induce high RAM usage with a linear dependence. $m$ should be carefully chosen, depending on the required resolution/decimation rate and the design complexity.  Excessive RAM usage may lead to long propagation time, thus to  potential data corruption.

For practical reasons, we also estimate the intercept point ($b$ in Fig.~\ref{fig:Omega-block-diagram}, with $b=\left<\phi_k\right>-a\left<t_k\right>$) leading to a more generic linear regression filter instantiation.
Table \ref{table:area} summarizes some implementation reports on the principal items that are BRAMs and DSP48, and their respective occupation for a Zynq-based system of the Zedboard type.

\begin{table}
\begin{center}
\caption{\label{table:area}}
\begin{tabular}{cccccc}
M & m & \#RAM & \%  & \#DSP48 & \%  \\
\hline\hline
32 & 1024 & 2/140 & 1.4 & 4/220 & 1.8 \\
32 & $2^{15}$ & 64/140 & 45 & 4/220 & 1.8 \\
64 & $2^{15}$ & 128/140 & 91.4 & 13/220 & 5.9 \\
\hline
\end{tabular}
\end{center}
\end{table}

\section{Discussion}

\subsection{Noise Rejection}
We summarize the noise response of the counters in the presence of a data series $\{\bx_k\}$ uniformly spaced by $\tau_0$, or equivalently sampled at the frequency $\nu_s=1/\tau_0$.  The measurement time is $\tau=m\tau_0$,
The noise samples $\sx_k$ associated to $\bx_k$ are statistically independent.

The analysis of the $\Pi$ and $\Lambda$ counters provided in this Section is based on \cite{rubiola2005,dawkins2007}, with a notation difference concerning the $\Lambda$ counter.

In \cite{rubiola2005,dawkins2007}, the measurement time spans over $2\tau$, and two contiguous measures are overlapped by $\tau$.  This choice is driven by the application to the modified Allan variance, having the same response $\text{mod}\sigma^2_\sy(\tau)=\frac{1}{2}\mathsf{D}_\sy^2\tau^2$ to a constant drift $\mathsf{D}_\sy$.

Oppositely, in this article use the same time interval $\tau$ for all the counters, and we leave the two sample variances to the companion article \cite{Vernotte-2015-arXiv--PVAR}.

\subsubsection{$\Pi$ counter}
The estimated fractional frequency is
\begin{align*}
\hat{\by} = \frac{\bx_{m-1}-\bx_0}{m\tau_0}  =
\frac{\bx_{m-1}-\bx_0}{\tau}.
\end{align*}
The associated variance is
\begin{align}
\mathbb{V}\{\hat{\by}\} =\frac{2\sigma_\sx^2}{m\tau_0^2}
=\frac{2\sigma_\sx^2}{\tau^2}, \label{eqn:Pi-variance}
\end{align}
independent of the sampling frequency.

\subsubsection{$\Lambda$ counter}
In the measurement time $\tau=m\tau_0$ we take $m/2$ almost-overlapped samples $\by_k$ measured by $\Pi$ counter a over a time $\tau/2$
\begin{align*}
\hat{\by}_k = \frac{\bx_{m/2+k}-\bx_k}{m\tau_0/2} =
\frac{\bx_{m/2+k}-\bx_k}{\tau/2}.
\end{align*}
The variance associated to each $\by_k$ is
\begin{align*}
\mathbb{V}\{\hat{\by}_k\} =\frac{2\sigma_\sx^2}{m\tau_0^2/4}.
\end{align*}
The $\Lambda$ estimator gives
$\hat{\by}=\frac{2}{m}\sum_{k=0}^{(m-1)/2}\by_k$.  Thus, the estimator variance is
\begin{gather}
\mathbb{V}\{\hat{\by}\}
\approx\frac{16\sigma _{\sx}^{2}}{m^3\tau_0^2}
=\frac{16\sigma _{\sx}^{2}}{\nu_s\tau^3}.
\label{eqn:Lambda-variance}
\end{gather}

\subsubsection{$\Omega$ counter}
The estimator variance \req{eqn:Omega-variance} can be rewritten in
terms of the measurement time $\tau=m\tau_0$ and the sampling
frequency $\nu_s=1/\tau_0$ as follows
\begin{gather}
\mathbb{V}\{\hat{\by}\}
\approx\frac{12\sigma _{\sx}^{2}}{m^3\tau_0^2}
=\frac{12\sigma _{\sx}^{2}}{\nu_s\tau^3}.
\label{eqn:Omega-variance-2}
\end{gather}

\subsubsection{Comparison}
Comparing the estimator variance $\mathbb{V}\{\hat{\by}\}$ as given by Eq.~\req{eqn:Pi-variance},
\req{eqn:Lambda-variance} and
\req{eqn:Omega-variance-2}
we notice that the $\Pi$ estimator features the poorest rejection to white PM noise, proportional to $1/\tau^2$.
The white PM noise rejection follows the $1/\tau^3$ law for both the $\Lambda$ and $\Omega$ estimators, but the $\omega$ estimator is more favorable by a factor of $3/4$, i.e., 1.25 dB\@.   However small, this benefit comes at no expense in term of the complexity of precision electronics.

The sampling frequency $\nu_s=1/\tau_0$ can be equal to the input frequency $\nu_0$ or an integer fraction of, depending on the speed of the interpolator and on the data transfer rate.  Values of 1--4 MHz are found in commercial equipment.

\subsection{Data Decimation}
We address the question of decimating a stream $\{\by_k(\tau)\}$ of contiguous data averaged over $\tau$, in order to get a new stream $\{\by_j(\tau)\}$ of contiguous data averaged over $n\tau$, preserving the statistical properties of the estimator.

\subsubsection{$\Pi$ counter}
Decimation is trivially done by averaging $n$ contiguous data with zero dead time
\begin{align*}
\hat{\by}(n\tau)=\frac{1}{n}\sum_{k=0}^{n-1}\hat{\by}_k(\tau).
\end{align*}

\subsubsection{$\Lambda$ counter}
Decimation can be done only in power of 2 ($n=2^N$), provided that contiguous samples are overlapped by exactly $\tau/2$.  In this case, we apply recursively the formula
\begin{align*}
\hat{\by}(2\tau)=
\frac{1}{4}\hat{\by}_{-1}(\tau) +
\frac{1}{2}\hat{\by}_{0}(\tau) +
\frac{1}{4}\hat{\by}_{1}(\tau).
\end{align*}
Notice that the measurement time $2\tau$ is given by the overlap between samples.

\subsubsection{$\Omega$ counter}
An exact decimation formula, as for the $\Pi$ and $\Lambda$ counters, does not exist.  The reason is that the exact parabolic shape has constant second derivative.  This is incompatible with the presence of the edges at $\pm\tau/2$ when more than one shifted instances of the $w(t)$ function are overlapped.

\subsection{Application to the two-sample variances}
The general form of the two-sample variance reads
\begin{equation}
\mathbb{V}\left\{\hat{\sy}\right\}=\frac{1}{2}\mathbb{E}\left\{\left(\hat{\sy}_1-\hat{\sy}_2\right)^2\right\}
\label{eqn:twosample}
\end{equation}
Using a stream of `contiguous' data $\{\hat{\sy}_k\}$ measured with a $\Pi$, a $\Lambda$ or a $\Omega$ estimator, Eq.~\req{eqn:twosample}, gives the AVAR, the MVAR or the PVAR\@.  Notice that in the case of the MVAR, `contiguous' means 50\% overlapped.

\section{Conclusions}
The same architecture (Fig.~\ref{fig:architecture}) is suitable to the $\Pi$, $\Lambda$ and $\Omega$ estimators, just by replacing the algorithm that processes the phase data.  A smart design can implement the three estimators with a some digital-hardware overhead.

The $\Pi$ estimator is the poorest, and mentioned only for completeness.  Nonetheless, it is still on the stage when extreme simplicity is required, or in niche applications like the measurement of the Allan variance.

In the presence of the same amount of white noise, the $\Omega$ estimator is similar to the $\Lambda$ estimator, but features smaller variance.
The difference is a factor of $3/4$ (1.25 dB).

The $\Lambda$ estimator is a good choice when a clean decimation of the output data is mandatory.  By contrast, the $\Omega$ estimator is superior for the detection of noise phenomena \cite{Vernotte-2015-arXiv--PVAR}

However may object that a noise reduction of 1.25 dB is not a big deal, the value our analysis is in showing that there is no room for further improvement.

If complexity is the major issue, the $\Lambda$ estimator fits, at the cost of 1.25 dB higher noise.  Oppositely, if the ultimate noise performance in the presence of wideband noise is an issue, the $\Omega$ estimator is the right answer.  The complexity of the LR is likely to be affordable on FPGA or SoC electronics.

\section*{Acknowledgements}
This work is supported by the ANR Programme d'Inve\-stis\-se\-ment d'Avenir in
progress at the Time and Frequency Departments of FEMTO-ST Institute and UTINAM (Oscillator IMP, First-TF and Refimeve+), and by the R\'{e}gion de Franche-Comt\'{e}.

\bibliography{omega.bib}

\begin{thebibliography}{10}
\providecommand{\url}[1]{#1}
\csname url@samestyle\endcsname
\providecommand{\newblock}{\relax}
\providecommand{\bibinfo}[2]{#2}
\providecommand{\BIBentrySTDinterwordspacing}{\spaceskip=0pt\relax}
\providecommand{\BIBentryALTinterwordstretchfactor}{4}
\providecommand{\BIBentryALTinterwordspacing}{\spaceskip=\fontdimen2\font plus
\BIBentryALTinterwordstretchfactor\fontdimen3\font minus
  \fontdimen4\font\relax}
\providecommand{\BIBforeignlanguage}[2]{{%
\expandafter\ifx\csname l@#1\endcsname\relax
\typeout{** WARNING: IEEEtran.bst: No hyphenation pattern has been}%
\typeout{** loaded for the language `#1'. Using the pattern for}%
\typeout{** the default language instead.}%
\else
\language=\csname l@#1\endcsname
\fi
#2}}
\providecommand{\BIBdecl}{\relax}
\BIBdecl

\bibitem{Dekatron}
Ericsson, ``{GC10B} {D}ekatron counter tube,'' The data sheet is no longer
  available, circa 1955, \mbox{S}ee https://en.wikipedia.org/wiki/Dekatron and
  http://www.electricstuff.co.uk/count.html\#deka.

\bibitem{Henzler-2010:Time-to-digital-converters}
S.~Henzler, \emph{Time-to-Digital Converters}.\hskip 1em plus 0.5em minus
  0.4em\relax Springer, 2010.

\bibitem{yoder:quantizers}
S.~Yoder, M.~Ismail, and W.~Khalil, \emph{{VCO}-Based Quantizers Using
  Frequency-to-Digital and Time-to-Digital Converters}.\hskip 1em plus 0.5em
  minus 0.4em\relax Springer, 2011.

\bibitem{Kalisz-2004-Metrologia--Counters}
J.~Kalisz, ``Review of methods for time interval measurements with picosecond
  resolution,'' \emph{Metrologia}, vol.~41, pp. 17--32, 2004.

\bibitem{Nutt-1968-RSI--Counters}
R.~Nutt, ``Digital time intervalometer,'' \emph{Rev.\ Sci.\ Instrum.}, vol.~39,
  no.~9, pp. 1342--1345, Sep. 1968.

\bibitem{Nutt-1976-Patent--Interpolator}
R.~Nutt, K.~Milam, and C.~W. Williams, ``Digital intervalometer,'' U. S. Patent
  no.\ 3,983,481, Sep.~28, 1976.

\bibitem{Cottini-Gatti-1956-NC--Vernier}
C.~Cottini and E.~Gatti, ``Millimicrosecond time analyzer,'' \emph{Nuovo
  Cimento}, vol.~4, p. 1550, 1956.

\bibitem{Cottini-Gatti-Giannelli-1956-NC--Vernier}
C.~Cottini, E.~Gatti, and G.~Giannelli, ``Millimicrosecond time analyzer,''
  \emph{Nuovo Cimento}, vol.~4, p. 156, 1956.

\bibitem{Lefevre-1957:Vernier-chronotron}
H.~W. Lefevre, \emph{The Vernier Chronotron}.\hskip 1em plus 0.5em minus
  0.4em\relax Chicago, IL, USA: Univ.\ of Michigan Library, Jan.~1, 1957.

\bibitem{Kindlmann-Sunderland-RSI-1966--Chronotron}
P.~J. Kindlmann and J.~Sunderland, ``Phase stabilized vernier chronotron,''
  \emph{Rev.\ Sci.\ Instrum.}, vol.~37, no.~4, pp. 445--452, Apr. 1966.

\bibitem{Chu-1989-HPJ--HP5371}
D.~C. Chu, ``Phase digitizing: A new method for capturing and analyzing
  spread-spectrum signals,'' \emph{Hewlett Packard J.}, vol.~40, no.~2, pp.
  28--35, Feb. 1989.

\bibitem{Acam}
ACAM, \texttt{http://www.acam.de/products/time-to-digital-converters/}.

\bibitem{brilliant}
Brilliant Instruments, \texttt{http://www.b-i-inc.com/}.

\bibitem{guidetech}
GuideTech \texttt{http://www.guidetech.com}.

\bibitem{Kramer2001}
G.~Kramer and K.~Klitsche, ``Multi-channel synchronous digital phase
  recorder,'' in \emph{Proceedings of the 2001 IEEE International Frequency
  Control Symposium}, Seattle (WA), USA, Jun. 2001, pp. 144--151.

\bibitem{K-K-Messtechnik}
K \& K Messtechnik GmbH, St.-Wendel-Str.\ 12, 38116 Braunschweig, Germany.

\bibitem{Keysight}
Keysight Technologies, \texttt{http://www.keysight.com}.

\bibitem{Maxim}
Maxim Integrated, \texttt{http://www.maximintegrated.com/}.

\bibitem{Spad}
SPAD Lab, \texttt{http://www.everyphotoncounts.com/index.php}.

\bibitem{texas}
Texas Instruments, \texttt{http://www.ti.com}.

\bibitem{rubiola2005}
E.~Rubiola, ``On the measurement of frequency and of its sample variance with
  high-resolution counters,'' \emph{Review of Scientific Instruments}, vol.~76,
  no. 054703, May 2005, article no. 054703.

\bibitem{dawkins2007}
S.~T. Dawkins, J.~J. McFerran, and A.~N. Luiten, ``Considerations on the
  measurement of the stability of oscillators with frequency counter,''
  \emph{\mbox{IEEE} Transactions on Ultrasonics, Ferroelectrics, and Frequency
  Control}, vol.~54, no.~5, pp. 918--925, May 2007.

\bibitem{allan1966}
D.~W. Allan, ``Statistics of atomic frequency standards,'' \emph{Proceedings of
  the IEEE}, vol.~54, no.~2, pp. 222--231, 1966.

\bibitem{Barnes-et-al-1971-TIM--Frequency-stability}
J.~A. Barnes, A.~R. Chi, L.~S. Cutler, D.~J. Healey, D.~B. Leeson, T.~E.
  Mc{G}unigal, J.~A. Mullen, Jr, W.~L. Smith, R.~L. Sydnor, R.~F.~C. Vessot,
  and G.~M.~R. Winkler, ``Characterization of frequency stability,'' \emph{IEEE
  Transact.\ on Instrum.\ Meas.}, vol.~20, no.~2, pp. 105--120, May 1971.

\bibitem{Snyder-1980-AO--Mod-Allan}
J.~J. Snyder, ``Algorithm for fast digital analysis of interference fringes,''
  \emph{Applied Optics}, vol.~19, no.~4, pp. 1223--1225, Apr. 1980.

\bibitem{Allan-Barnes-1981-FCS--Mod-Allan}
D.~W. Allan and J.~A. Barnes, ``A modified ``{A}llan variance'' with increased
  oscillator characterization ability,'' in \emph{Proc.\ 35 IFCS},
  Ft.~Monmouth, NJ, May 1981, pp. 470--474.

\bibitem{Lesage-Ayi-1984-IM--Mod-Allan}
P.~Lesage and T.~Ayi, ``Characterization of frequency stability: Analysis of
  the modified {A}llan variance and properties of its estimate,'' \emph{IEEE
  Transact.\ on Instrum.\ Meas.}, vol.~33, no.~4, pp. 332--336, Dec. 1984.

\bibitem{Johansson2005fcs-counters}
S.~Johansson, ``New frequency counting principle improves resolution,'' in
  \emph{Proc.\ Freq.\ Control Symp.}, Vancouver, BC, Canada, Aug.~29--31, 2005,
  pp. 628--635.

\bibitem{Benkler2015arxiv}
\BIBentryALTinterwordspacing
E.~Benkler, C.~Lisdat, and U.~Sterr, ``On the relation between uncertainties of
  weighted frequency averages and the various types of \mbox{Allan}
  deviations,'' arXiv:1504.00466, April 2015. [Online]. Available:
  \url{http://arxiv.org/pdf/1504.00466v3.pdf}
\BIBentrySTDinterwordspacing

\bibitem{Greenhall-1989-UFFC--Picket-fence}
C.~A. Greenhall, ``A method for using time interval counters to measure
  frequency stability,'' \emph{IEEE Trans.\ Ultras.\ Ferroelec.\ Freq.\
  Contr.}, vol.~36, no.~5, pp. 478--480, Sep. 1989.

\bibitem{Greenhall-1997-IM--Picket-fence}
------, ``The third-difference approach to modified {A}llan variance,''
  \emph{IEEE Transact.\ on Instrum.\ Meas.}, vol.~46, no.~3, pp. 696--703, Jun.
  1997.

\bibitem{Vernotte-2015-Metrologia}
\BIBentryALTinterwordspacing
F.~Vernotte and E.~Lantz, ``Metrology and 1/ f noise: linear regressions and
  confidence intervals in flicker noise context,'' \emph{Metrologia}, vol.~52,
  no.~2, pp. 222--237, 2015. [Online]. Available:
  \url{http://stacks.iop.org/0026-1394/52/i=2/a=222}
\BIBentrySTDinterwordspacing

\bibitem{Altera}
Altera Corp., USA, \texttt{http://altera.com/}.

\bibitem{Calosso-2014-IFCS--Digital}
C.~E. Calosso and E.~Rubiola, ``Phase noise and jitter in digital electronic
  components (invited article),'' in \emph{Proc.\ Int'l Freq.\ Control Symp.},
  Taipei, Taiwan, May~19--22, 2014, pp. 532--534.

\bibitem{Xilinx}
Xilinx, \texttt{http://www.xilinx.com}.

\bibitem{Rubiola-1992-FCS-Trigger}
E.~Rubiola, A.~Del~Casale, and A.~De~Marchi, ``Noise induced time interval
  biases,'' in \emph{Proc.\ Int'l Freq.\ Control Symp.}, Hershey, PA, USA,
  May~27--29, 1992, pp. 265--269.

\bibitem{Vernotte-2015-arXiv--PVAR}
F.~Vernotte, M.~Lenczner, P.-Y. Bourgeois, and E.~Rubiola, ``The parabolic
  variance (pvar), a wavelet variance based on least-square fit,''
  arXiv:1506.00687 [physics.data-an], Jun.~1, 2015, \mbox{S}ubmitted to IEEE
  Transact UFFC, Special Issue on the 50th Anniversary of the Allan Variance.

\end{thebibliography}
\end{document}